\documentclass[twocolumn,amsmath,amssymb,prl,footinbib,10pt,superscriptaddress]{revtex4-2}
\pdfoutput=1
\usepackage{ graphicx, float,amsmath, amsmath, amssymb}
\usepackage[export]{adjustbox}

\def\be{\begin{align}}
\def\ee{\end{align}}
\def\bea{\begin{eqnarray}}
\def\eea{\end{eqnarray}}
\def\bse{\begin{subequations}}
\def\ese{\end{subequations}}

\usepackage[breaklinks, colorlinks,linkcolor=black, citecolor=black,urlcolor = black]{hyperref}
\usepackage[labelsep=period]{caption}
\begin{document}

\title{Gauge fields and quantum entanglement}

\author{Jakub Mielczarek}
\email{jakub.mielczarek@uj.edu.pl}
\author{Tomasz Trze\'sniewski}
\email{t.trzesniewski@uj.edu.pl}
\affiliation{Institute of Theoretical Physics, Jagiellonian University, ul. {\L}ojasiewicza 11, 30-348 Krak\'{o}w, Poland}

\date{\today}

\begin{abstract}
The purpose of this letter is to explore the relation between gauge fields, which are at the base 
of our understanding of fundamental interactions, and the quantum entanglement. To this end, 
we investigate the case of ${\rm SU}(2)$ gauge fields. It is first argued that holonomies of 
the ${\rm SU}(2)$ gauge fields are naturally associated with maximally entangled two-particle states. 
Then, we provide some evidence that the notion of such gauge fields can be deduced from the 
transformation properties of maximally entangled two-particle states. This new insight unveils a 
possible relation between gauge fields and spin systems, as well as contributes to understanding 
of the relation between tensor networks (such as MERA) and spin network states considered in 
loop quantum gravity. In consequence, our results turn out to be relevant in the context of the 
emerging Entanglement/Gravity duality.
\end{abstract}

\maketitle

\section{Introduction} 

Entanglement is usually considered to be a solely quantum mechanical 
phenomenon. However, it has been broadly argued in the recent years that classical 
geometry provides a description of the structure of entanglement in some 
quantum systems. One way to uncover this relation is by using tensor networks (graphs constructed from 
contracted tensors), which represent wave functions of certain many-body 
quantum systems \cite{Biamonte:2017dgr}. In particular, entanglement renormalization of a 
many-body system, performed in terms of tensor networks, leads to anti-de Sitter (AdS) 
spacetime geometry \cite{Swingle:2009bg}, which plays a central role in the AdS/CFT correspondence 
\cite{Maldacena:1997re}. While the tensor networks are discrete objects, their continuous 
limits are possible to define \cite{Tilloy:2018gvo}. This opens a possibility that continuous 
space-time may be considered as an approximation to a discrete tensor network representing
the entanglement structure of a certain discrete (e.g. spin) system. In the holographic
language, the quantum system under consideration represents the boundary and 
the bulk geometry is represented by the entanglement structure between quantum 
degrees of freedom at the boundary. Further evidence for such a viewpoint comes 
from considerations of the entanglement entropy \cite{Ryu:2006bv} and quantum 
complexity \cite{Brown:2017jil}. 

While the bulk geometry in the above context is considered to be classical,
theories such as loop quantum gravity (LQG) \cite{Ashtekar:2004eh, RovelliVidotto} 
provide the quantum viewpoint on the nature of spacetime. The question that arises 
is whether the quantum description of spacetime is consistent with the (classical) geometry 
describing the entanglement of a many-body quantum system or a field theory. 
However, at least in the LQG approach, a possibility to merge the two viewpoints 
occurs. Namely, the spatial geometry in LQG is described by a graph called the 
\emph{spin network}, which is an object built from the holonomies of Ashtekar 
$\mathfrak{su}(2)$ gauge fields \cite{Ashtekar:1986yd}. A state of the spin network 
is constructed as a contraction of the holonomies, so that the obtained function is 
invariant under local gauge transformations (i.e. it is annihilated by the Gauss constraint). 
 
It was argued in \cite{Han:2016xmb} that coarse-graining of spin network states 
leads to the emergence of a tensor network at the boundary of space. The underlying 
reason is that when open spin networks are considered, the endpoints of open links, 
which explicitly break the local gauge invariance, can be associated with the (boundary) 
degrees of freedom. On the other hand, (superpositions of) spin networks themselves 
can be treated as tensor networks \cite{Singh:2010ty} (see also \cite{Chirco:2018gy} for 
a generalization of this relation to the group field theory framework). The links (holonomies) 
of spin networks are, therefore, expected to be inevitably related to entanglement 
\cite{Mielczarek:2018jsh} (in particular, maximally entangled states impose specific 
gluing conditions on spatial polyhedra associated with spin network vertices \cite{Baytas:2018wjd}). 

The purpose of this letter is to explore these relationships further. However, let us stress that 
while our inspiration comes from LQG, the analysis will not be restricted to this theory -- it 
concerns a general relation between holonomies and (maximally) entangled states.

\section{Gauge fields} 

Let us begin our considerations from a $\mathfrak{su}(2)$ gauge field $A^i_a$, with 
algebra indices $i = 1,2,3$ and spatial indices $a = 1,2,3$. The field is characterized by 
the connection 1-form $A = A^i_a \tau_i dx^a$, where $\tau_i$ generate the $\mathfrak{su}(2)$ 
algebra $[\tau_i,\tau_j] = \epsilon_{ijk} \tau_k$. In the fundamental representation, $\tau_i$ 
are related to Pauli matrices via the relation $\tau_i = -\frac{i}{2} \sigma_i$.

The essential feature of gauge field theories is invariance with respect to 
local gauge transformations: 
\begin{align}
A_a \rightarrow A'_a = U^\dagger A_a U + U^\dagger \partial_a U\,,
\label{GaugeTransformA}
\end{align}
where $U$ is a certain unitary matrix. While the form of a transformation 
of the gauge field rather does not tell us too much directly, there are other objects 
that allow us to look at the gauge transformation from slightly different perspective. 
An example of such an object on which we are going to focus our attention is
the holonomy of a gauge field, which is an ${\rm SU}(2)$ element defined as follows: 
\begin{align}
h_e[A] := \mathcal{P} \exp\int_e A\,,  
\label{HolonomyofA}
\end{align}
where $e$ is a path $e: [0,1] \rightarrow \Sigma$, intermediating between
the source $e(0) = s$ and target $e(1) = t$ points on a 
spatial hypersurface $\Sigma$, and $\mathcal{P}$ denotes the path ordering. The 
holonomy is clearly a non-local object, which under the transformation (\ref{GaugeTransformA})
transforms as
\begin{align}
h_e[A] \rightarrow h'_e[A] & = U^\dagger(e(0))\, h_e[A]\, U(e(1)) \nonumber\\
&= U_s^{\dagger} h_e[A] U_t\,,
\label{HolonomyGauge}
\end{align}
where for further convenience we defined $U_s := U(e(0))$ and $U_t := U(e(1))$. 
The gauge transformation acts at the holonomy only at the endpoints. 
This property is widely used in the construction of lattice gauge theories, especially 
by introducing gauge-invariant Wilson loops $W_e[A] := \text{tr}(h_e[A])$, 
where $e$ is then a closed path.

\section{Unitary map} 

In the gauge theory context, holonomies are parallel-transported 
elements of the gauge group. However, they can also be treated as isomorphisms 
between certain linear spaces. In order to see this explicitly, let us consider the 
holonomy (\ref{HolonomyofA}) in the case of the fundamental representation of 
${\rm SU}(2)$, i.e. spin-1/2. Then, the holonomy is given by a $2\times 2$ ${\rm SU}(2)$ 
matrix, which belongs to the automorphism group of $\mathbb{C}^2$ (i.e. the space 
of non-relativistic spinors). 

We will now make the qualitative jump to quantum mechanics, using the fact that 
$\mathbb{C}^2$ equipped with the natural scalar product becomes the Hilbert space of a 
\emph{qubit} system. (Pure) physical states correspond to rays in $\mathbb{C}^2$, i.e. 
elements of $\mathbb{CP}^1$, which can be represented as the (${\rm SU}(2)$-invariant) unit 
sphere. Thus, the holonomy is an isomorphism between the two 2-dimensional (projective) 
Hilbert spaces $\mathcal{H}_s = \text{span}\left\{| 0 \rangle_s, 
| 1  \rangle_s\right\}$ and $\mathcal{H}_t = \text{span}\left\{| 0 \rangle_t, 
| 1 \rangle_t\right\}$, in which we choose orthonormal bases, i.e. ${_{s}}\langle I 
| J \rangle_s = \delta_{IJ}$ and ${_{t}}\langle I | J \rangle_t = \delta_{IJ}$, where 
$I,J = 0,1$. $\mathcal{H}_s$ and $\mathcal{H}_t$ are assumed to be associated with two 
different points in space. Since now the holonomy (\ref{HolonomyofA}) is represented 
by a unitary matrix, the corresponding map (isomorphism) $h$ is unitary as well. Let us 
now see whether a transformation (\ref{HolonomyGauge}) of such a map arises in some 
natural way in the quantum-mechanical context. 

Employing the bases of the source and target Hilbert spaces 
($\mathcal{H}_s$ and $\mathcal{H}_t$), it is convenient to express an arbitrary unitary
map between $\mathcal{H}_s$ and $\mathcal{H}_t$ as
\begin{align}
h = h_{IJ} | I \rangle_s {_t}\langle J | \in \mathcal{H}_s \otimes \mathcal{H}_t^*\,, \label{holmat}
\end{align}
where $h_{IJ}$ are matrix elements of $h$ and $\mathcal{H}_t^*$ is the space dual to 
$\mathcal{H}_t$. The action of this map can be either left-handed or right-handed, 
so that $h_L : \mathcal{H}_s^* \rightarrow \mathcal{H}_t^*$ and 
$h_R : \mathcal{H}_t \rightarrow \mathcal{H}_s$. Analogously, the Hermitian 
conjugation of $h$, $h^\dagger = h_{IJ}^* | J \rangle_t {_s}\langle I | \in \mathcal{H}_t 
\otimes \mathcal{H}_s^*$, acts as $h^\dagger_L : \mathcal{H}_t^* 
\rightarrow \mathcal{H}_s^*$ or $h^\dagger_R : \mathcal{H}_s \rightarrow \mathcal{H}_t\,$. 
For example, a basis state ${_s}\langle K | \in \mathcal{H}_s^*$ at a point $s$ is mapped 
according to ${_s}\langle K |\, h = h_{IJ} {_s}\langle K | I \rangle_s {_t}\langle J | = 
h_{KJ} {_t}\langle J | \in \mathcal{H}_t^*\,$ into a certain state at a point $t$. 

\begin{figure}[ht!]
\centering
\includegraphics[scale=0.3]{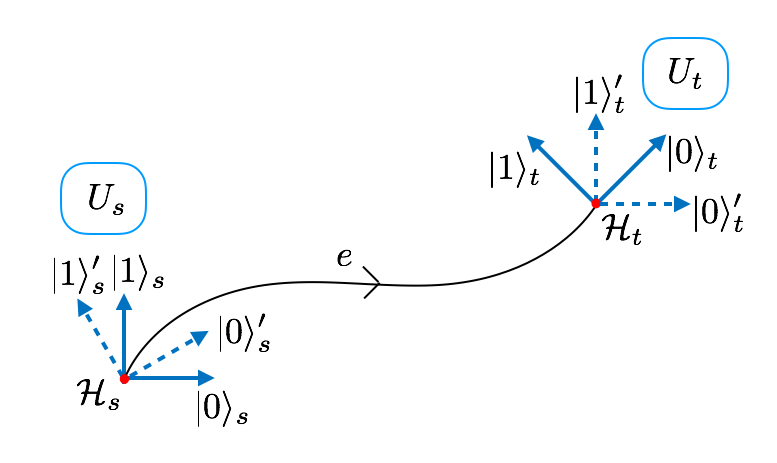}
\caption{Graphical representation of the change of bases of the source and the target
Hilbert spaces, associated with the space points $s$ and $t$ respectively.}
\label{Bases}
\end{figure}

Let us now proceed to the crucial point. The choice of basis in both the source 
and target Hilbert space is completely arbitrary. Therefore, we can ask how the map (\ref{holmat}) behaves 
under the action of unitary transformations that change these bases. The transformations at the source 
and target can in general be different and we will distinguish them by using the 
indices $s$ and $t$. A unitary transformation of a basis state $| I \rangle$ can be 
written as $| I \rangle' = U | I \rangle$ or -- in terms of matrix elements $U_{IJ}$ of $U$ -- as 
$| I \rangle' = U_{JI} | J \rangle\,$. The transformation of bases at the source $s$ and 
the target $t$ has been pictorially presented in Fig. \ref{Bases}.

\begin{figure}[ht!]
\centering
\includegraphics[scale=0.25]{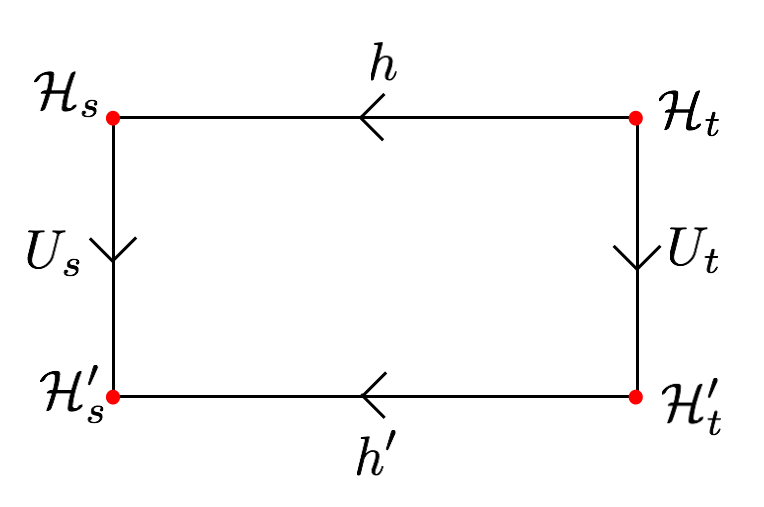}
\caption{Graphical representation of the transformation rule of holonomy
under the change of bases.}
\label{Map}
\end{figure}

Applying such transformations to both source and target bases in (\ref{holmat}), we find that
\begin{align}
h_{IJ} | I \rangle_s {_t}\langle J | &= h'_{IJ} | I \rangle_s' {_t}\langle J |' \nonumber\\ 
&= U_{s,KI} h'_{IJ} U_{t,JL}^{\dagger}| K \rangle_s {_t}\langle L |\,,
\label{MapATransform}
\end{align}
which leads to the following transformation rule:
\begin{align}
h \rightarrow h' = U_s^\dagger h U_t\,,
\label{MapTransform}
\end{align}
which is graphically represented in Fig. \ref{Map}.

It is clear that the above change of $h$ under unitary transformations 
$U_s$ and $U_t$ in the source and target spaces is formally equivalent to the 
action of a ${\rm SU}(2)$ gauge transformation. At least in this sense, a map (\ref{holmat}) 
shares properties of a holonomy along the curve connecting $s$ and $t$. 
Analysis in the next section will underscore important consequences of this simple fact.
However, if and how a map (\ref{holmat}) may inherit any information about the curve is a much 
more involved issue, which we do not consider here. In the next sections of this letter we 
will provide some evidence for the conjecture that ${\rm SU}(2)$ gauge transformations 
reflect the invariance of physics of the corresponding quantum system under unitary 
transformations of both bases.

\section{Anti-linear map}

Since the holonomy is associated with a pair of Hilbert spaces 
$\mathcal{H}_s$ and $\mathcal{H}_t$, it is natural to ask whether there are some 
interesting states belonging to $\mathcal{H}_s \otimes \mathcal{H}_t$ that can be defined using $h$. Following Refs. 
\cite{Czech:2018kvg,Czech:2019vih}, let us consider a state 
\begin{align}
| \Psi \rangle := \frac{1}{\sqrt{2}} h^*_{IJ} | I \rangle_s | J \rangle_t \in \mathcal{H}_s \otimes \mathcal{H}_t\,,
\label{PsiQubit}
\end{align}
where $h_{IJ}$ are matrix elements of the ${\rm SU}(2)$ holonomy. 

The fact that coefficients of the state (\ref{PsiQubit}) are given by components 
of a special unitary matrix has profound consequences. Namely, this implies 
that the state (\ref{PsiQubit}) belongs to the class of \emph{maximally entangled states}, 
for which the reduced density matrix is diagonal (as it is well known in quantum information 
theory; the space of such states for a 2-qubit system is actually ${\rm SU}(2)/\mathbb{Z}_2$ 
\cite{Bengtsson:2017gs}). Explicitly, the density matrix associated with the state (\ref{PsiQubit}) 
has the form $\hat{\rho} = | \Psi \rangle \langle \Psi | = 
\frac{h_{IJ}^* h_{KL}}{2} \left(| I \rangle_s {_s}\langle K |) (| J \rangle_t {_t}\langle L |\right)$.
In consequence, the reduced density matrix $\hat{\rho}_s := \text{tr}_t(\hat{\rho}) = 
\frac{h_{IJ}^* h^T_{JK}}{2} \left(| I \rangle_s {_s}\langle K |\right) = \frac{1}{2} \hat{I}$,
where unitarity of the matrix $[h_{IJ}]$ has been used. The analogous formula can be found 
for $\hat{\rho}_t := \text{tr}_s(\hat{\rho})$. 

While distinguishing the state (\ref{PsiQubit}) might seem arbitrary, it has been 
shown in \cite{Czech:2019vih} that (\ref{PsiQubit}) can be used to define the holonomy 
map between the (dual) source space and target space. Namely, the idea is to consider 
an appropriate anti-linear map (each such a map can be decomposed into a linear map 
and the complex conjugation $C$; in the physical context, $C$ can arise e.g. via the CPT 
transformation), which in our notation acts on basis states as:  
\begin{align}
\mathcal{H}_s^* \ni {_s}\langle I | \rightarrow \left[\sqrt{2}\, | \Psi \rangle \circ C\right] \left({_s}\langle I |\right)
= h^*_{IJ} | J \rangle_t\, \in \mathcal{H}_t\,.
\label{AntiMap}
\end{align}
where the state $| \Psi \rangle$ is given by (\ref{PsiQubit}). Applying the operation $Q \equiv \sqrt{2}\, | \Psi \rangle \circ C$ 
to an arbitrary state $c_I \langle I |_s$, we have:
$\mathcal{H}_s^* \ni c_I \langle I |_s \rightarrow Q \left(c_I \langle I |_s\right)
= h^*_{IJ} c_I^* | J \rangle_t\, \in \mathcal{H}_t\,$.

Furthermore, performing the analogous change of bases as in (\ref{MapATransform}) and using the map (\ref{AntiMap}), we obtain 
\begin{align}
{_s}\langle I |' &\rightarrow \left[\sqrt{2}\, | \Psi \rangle \circ C\right] \left({_s}\langle I |'\right) = 
\sqrt{2} \left(U^*_{s,JI}\right)^* {_s}\langle J | \Psi \rangle \nonumber\\ 
&= U_{s,IJ}^T h^*_{JL} | L \rangle_t = U_{s,IJ}^T h^*_{JL} U^*_{t,LM} | M \rangle_t'\,.
\label{BasisTransform1}
\end{align}
The transformation rule of coefficients $h_{JL}$ under the basis change is therefore:
$h_{JM} \rightarrow h_{JM}' = U_{s,IJ}^\dagger h_{JL} U_{t,LM}\,$, which is consistent 
with the gauge transformation of the holonomy (cf. (\ref{HolonomyGauge}) and (\ref{MapTransform})).

\section{Spatial entanglement from holonomies} 

The discussion presented so far indicates the existence of a non-trivial relation between holonomies 
of gauge fields and maximally entangled quantum states. Ref. \cite{Czech:2018kvg} has introduced 
the concept of \emph{entanglement holonomies}, used to define the quantum version of parallel transport 
between reference frames. This is achieved by the quantum teleportation of an auxilliary state via a 
maximally entangled state shared by the frames. Here, we would like to present a different perspective. 

Let us consider the following situation. Initially, we have two qubits at the source
(the analogous reasoning can be done for the target) and a total state of the 
system can be written as:
\begin{align}
| \phi_s \rangle = S_{KL} | K \rangle_s | L \rangle_s \in \mathcal{H}_s \otimes \mathcal{H}_s\,,
\end{align}
where $S_{KL}$ are some coefficients. Next, we would like to map one of the qubits from the
source to the target. This means that we take one of the qubits and move it (in 
the gauge field $A$) from $s$ to another space point $t$. If we choose the second qubit to 
be the one that is moved, the corresponding unitary map is:
\begin{align}
\mathbb{I} \otimes h^{\dagger}_R : \mathcal{H}_s \otimes \mathcal{H}_s \rightarrow  
\mathcal{H}_s \otimes \mathcal{H}_t
\end{align}
($h^\dagger_R$ is the right-hand action of the Hermitian conjugation of the 
holonomy $h$), which leads to  
\begin{align}
| \phi_s \rangle &\rightarrow S_{KL} | K \rangle_s  h^{\dagger}| L \rangle_s 
= S_{KL} h_{LJ}^* | K \rangle_s | J \rangle_t \nonumber\\
&\equiv C_{KJ} | K \rangle_s | J \rangle_t =: | \phi_{st} \rangle \in \mathcal{H}_s \otimes \mathcal{H}_t\,.
\end{align}
Coefficients of the obtained state $| \phi_{st} \rangle$ are related to coefficients 
of $| \phi_s \rangle $ through the relation $C_{KJ} = S_{KL} h_{LJ}^*$. If the state 
$| \phi_{st} \rangle$ is equivalent to (\ref{PsiQubit}), then $C_{KJ} = \frac{1}{\sqrt{2}} h^*_{KJ}$ 
and in consequence $S_{KL} = \frac{1}{\sqrt{2}} \delta_{KL}$. In such a case, $| \phi_{s} \rangle$ 
is a Bell state: $| \phi_{s} \rangle = \frac{1}{\sqrt{2}} \left( | 0 \rangle_s | 0 \rangle_s 
+ | 1 \rangle_s | 1 \rangle_s \right) =: | \Phi^+ \rangle\,$. 

Therefore, a maximally entangled state given by Eq. (\ref{PsiQubit}) can be 
obtained via the following map: 
\begin{align}
| \Psi \rangle = \left(\mathbb{I} \otimes h^\dagger_R\right) | \Phi^+ \rangle\,. 
\label{PsiPsiPlus}
\end{align}
A maximally entangled state of two qubits at two space points $s$ and $t$
can be considered the result of the holonomy acting on one of the two qubits initially 
located at the same space point. It is, however, necessary that a 2-qubit state is 
initially the $| \Phi^+ \rangle$ Bell state. Therefore, the above construction requires 
the initial entanglement of qubits, which can be generated 
from a non-entangled state as e.g. $| \Phi^+ \rangle = \text{CNOT} \circ 
(\text{H} \otimes \mathbb{I})\, | 0 \rangle_s | 0 \rangle_s$, where $\text{CNOT}$ 
and $\text{H}$ denote respectively the controlled-NOT and Hadamard quantum gates. 

\section{Gauge transformation from entanglement} 

We have shown that maximally entangled 2-particle quantum states emerge 
from considerations of holonomies of gauge fields. One may now ask whether 
this relation could also work in the opposite direction. As an inspiration, it is 
worth to mention that systems of spins (which, in the quantum informatic context, 
describe qubits) have been widely studied as the infrared approximation of 
${\rm SU}(2)$ Yang-Mills theory \cite{Faddeev:1999}. The first thing we recall 
here is that any maximally entangled 2-qubit state is characterized by 
a ${\rm SU}(2)$ matrix (up to the $\mathbb{Z}_2$ degeneracy, i.e. the overall sign). 

For simplicity, and to give an explicit example of the employed procedure, let us 
consider the following 2-qubit singlet state:
\begin{align}
|\Psi\rangle = | \Phi^- \rangle := \frac{1}{\sqrt{2}} \left( | 0 \rangle_s | 1 \rangle_t
- | 1 \rangle_s | 0 \rangle_t \right).
\end{align}
Comparing this state with (\ref{PsiQubit}), we observe that its coefficients correspond 
to the ${\rm SU}(2)$ matrix:
\begin{align}
h = \left( \begin{array}{cc} 0  & 1 \\ -1 & 0 \end{array} \right) = i \sigma_y 
= e^{i \frac{\pi}{2} \sigma_y}.
\label{HoloExample}
\end{align}
$h$ is again interpreted as a unitary map (holonomy) between the $\mathcal{H}_s$ and $\mathcal{H}_t$ 
Hilbert spaces, which transforms under the change of their bases according to (\ref{MapTransform}). 
In general, we have a three-parameter family of unitary basis transformations. Here, for simplicity, 
let us consider a one-parameter family of rotations:
\begin{align}
U(\theta) = e^{-i \theta \sigma_y} = \left(\begin{array}{cc} \cos\theta & -\sin\theta \\ \sin\theta & \cos\theta \end{array} \right).
\label{UExample}
\end{align}
The parameter $\theta$ is assumed to change smoothly between the values $\theta_s := \theta (e(0) = s)$ 
and $\theta_t := \theta (e(1) = t)$, corresponding to $U_s$ and $U_t$, as well as to satisfy the limiting condition $\lim_{t \to s} \theta_t = \theta_s$ (see also the discussion around (\ref{ContTrans})). Under such a transformation
of both bases, the map (\ref{HoloExample}) transforms as follows:
\begin{align}
h \rightarrow h' &= U_s^\dagger h U_t = e^{i \frac{\pi}{2} \sigma_y - i (\theta_t - \theta_s) \sigma_y} \nonumber\\
&= h e^{-i \sigma_y \int_e \partial_a\theta\, dx^a}.
\label{ExhTransform}
\end{align}
One can, therefore, conclude that the matrix (\ref{HoloExample}) can be written as exponentiation 
of a certain integral, such that its integrand under the change of bases transforms as 
\begin{align}
\text{integrand} \rightarrow \text{integrand} - i \partial_a\theta\,,
\label{U1gauge}
\end{align}
which has the same form as a ${\rm U}(1)$ gauge transformation. The integrand is what we can call 
a gauge field. This is how the concepts of a gauge symmetry and gauge field
can be deduced from considerations on maximally entangled states. 

In the considered example, the apparent ${\rm U}(1)$ gauge symmetry is actually 
a remnant of the ${\rm SU}(2)$ gauge symmetry restricted to the case of a 
one-parameter family of transformations. This can be seen by substituting 
(\ref{UExample}) into the transformation rule (\ref{GaugeTransformA}).
From the form of (\ref{HoloExample}), we see that the only non-vanishing 
component of the gauge field is $A^2_a$ (which is contracted with 
$\sigma_2 = \sigma_y$). In consequence, the gauge transformation reduces to: 
\begin{align}
A^2_a \rightarrow A^{2'}_a = A^2_a - i \partial_a \theta
\end{align}
and, making a comparison with (\ref{ExhTransform}), one can conclude that 
$\int_e A^2_a dx^a = -\pi$.

The above observations should be confirmed by analysis of the general 
case but such considerations will be rather computationally involving and 
are beyond the scope of this letter. Therefore, instead of this, let us make 
some crucial remarks and later provide another simple example. A general 
${\rm SU}(2)$ transformation of basis of either ${\cal H}_s$ or ${\cal H}_t$ 
can be obviously expressed in the form:
\begin{align}
U(\vec{a}) = e^{i \hat{a} \cdot \vec{\sigma}} =  \cos a\, \mathbb{I} + i \sin a\, (\hat{n} \cdot \vec{\sigma})\,,
\end{align} 
parametrized by a vector $\vec{a} = a\, \hat{n}$, where $a \equiv \sqrt{\vec{a} \cdot \vec{a}}$ 
and $\hat{n}$ is a versor. In order to be able to obtain the correspondence between a maximally 
entangled 2-particle state and a holonomy, we first have to restrict the allowed transformations 
to pairs of $U_s = U_s(\vec a_s)$ and $U_t = U_t(\vec a_t)$ that satisfy the conditions
\begin{align}\label{ContTrans}
\lim_{t \to s} U_t = U_s\ \Leftrightarrow\ \lim_{t \to s} \vec a_t = \vec a_s\,, \nonumber\\ 
\forall a: (\vec a_{s/t})^a \in {\cal C}^1(\Sigma)\,,
\end{align}
which means in particular that $U_s$ and $U_t$ are not independent. The next step 
is to make an extension to families of transformations $U(\vec a(\tau))$ acting 
in a family of Hilbert spaces ${\cal H}_\tau$, such that $U(\vec a(0)) = U_s$ and $U(\vec a(1)) = U_t$, 
as a generalization of (\ref{UExample}). Subsequently, one should check whether $h$ is indeed equivalent to a 
holonomy of the gauge field.

As an illustration, we will now consider the second example, given by the state
\begin{align}
|\Psi\rangle = | \Phi^+ \rangle := \frac{1}{\sqrt{2}} \left(| 0 \rangle_s | 0 \rangle_t + | 1 \rangle_s | 1 \rangle_t \right),
\end{align}
for which holonomy is simply an identity operator, $h = \mathbb{I}$. If we choose the vectors 
$\vec{a}_s$, $\vec{a}_t$ at the source and target, respectively, the transformation 
(\ref{MapTransform}) takes the form:
\begin{align}
h \rightarrow h' &= U_s^\dagger U_t = e^{i \alpha \hat{N} \cdot \vec{\sigma}},
\label{ExTransform2}
\end{align}
where
\begin{align}
\cos\alpha &= \cos a_s \cos a_t + \sin a_s \sin a_t\, \hat{n}_s \cdot \hat{n}_t\,, \label{cosalpha} \\
\sin\alpha\, \hat{N} &= \sin a_s \sin a_t\, (\hat{n}_s \times \hat{n}_t) + \cos a_s \sin a_t\, \hat{n}_t \nonumber\\
&- \sin a_s \cos a_t\, \hat{n}_s\,. \label{vecN}
\end{align}
While in general, $\hat{n}_s$ and $\hat{n}_t$ are different, the case with $\hat{n}_s = \hat{n}_t =: \hat{n}$ 
is especially illustrative and corresponds to rotations around a fixed axis. Then, (\ref{cosalpha}) 
reduces to $\cos{\alpha} = \cos(a_t - a_s)$ and (\ref{vecN}) becomes $\hat{N} = \hat{n}$. In consequence, 
the transformation (\ref{ExTransform2}) simplifies to $h \rightarrow h' = U_s^\dagger U_t = e^{i (\vec{a}_t - \vec{a}_s) \cdot \vec{\sigma}}$ 
and it can be rewritten in the form:
\begin{align}
e^{i (\vec{a}_t - \vec{a}_s) \cdot \vec{\sigma}} = e^{i (\int_e \partial _a \vec{a}\, dx^a) \cdot \vec{\sigma}}.
\label{ExTransform3}
\end{align}
 
The last expression has been inferred under the assumption that $U_s$, $U_t$ 
are the initial and final elements of a family of operators constructed by continuously 
changing the parameter vector $\vec{a}$ of the group element, in accordance with 
the postulates (\ref{ContTrans}) and below. This allows to associate a curve $e$ in 
space to any such family. It turns out that (\ref{ExTransform3}) agrees with what is 
expected from considerations of gauge transformations (\ref{GaugeTransformA}). 
Since for the example $h = \mathbb{I}$ we have $A_a = 0$, the pure gauge obtained by 
applying to $h$ a gauge transformation (\ref{GaugeTransformA}) leads to the following contribution 
to the exponent of the holonomy: 
\begin{align}
&\int_e U^{\dagger} \partial_a U dx^a = \nonumber\\
&= i \vec{\sigma} \cdot \int_e (\cos a\, \partial_a \tilde{n} - \tilde{n}\, \partial_a \cos a 
+ \tilde{n} \times \partial_a \tilde{n})\, dx^a,   
\end{align}
where $\tilde n := \sin a\, \hat{n}$. It is straightforward to check that the above formula 
gives the exponent in (\ref{ExTransform3}) for $\hat{n} =$ const.

\section{Spin networks and tensor networks} 

We are now ready to briefly discuss application of the prior discussion to spin 
networks and tensor networks. Since full discussion of the issue goes beyond 
the scope of this letter, we will only refer to some essential observations.  
 
Firstly, spin networks (without open links), which span the Hilbert space of LQG, 
by virtue of the loop transformation can be expressed as sums of products of 
the Wilson loops for the case of the fundamental representation of ${\rm SU}(2)$ \cite{Rovelli:1989za}. 
As it is clear from the definition (\ref{PsiQubit}), components 
of a holonomy are obtained by projecting the state $|\Psi \rangle$ on the 
basis states:
\begin{align}
h_{IJ} = \sqrt{2} \left( ({_s}\langle I | {_t}\langle J|)|\Psi \rangle \right)^* 
= \sqrt{2} \langle \Psi | (| I \rangle_s | J \rangle_t)
\end{align}
and a Wilson loop (for which $s = t$) can be expressed as
\begin{align}
W_e[h] = \sum_I h_{II} = \sqrt{2} \sum_I \langle \Psi | (| I \rangle | I \rangle)\,.
\end{align}
Therefore, the Wilson loop is a certain amplitude associated with the state $|\Psi \rangle$.
In consequence, the full spin network state can be expressed in terms of amplitudes of its 
constituent loops described by the corresponding states $|\Psi \rangle$.

\begin{figure}[h!]
\centering
\includegraphics[scale=0.3]{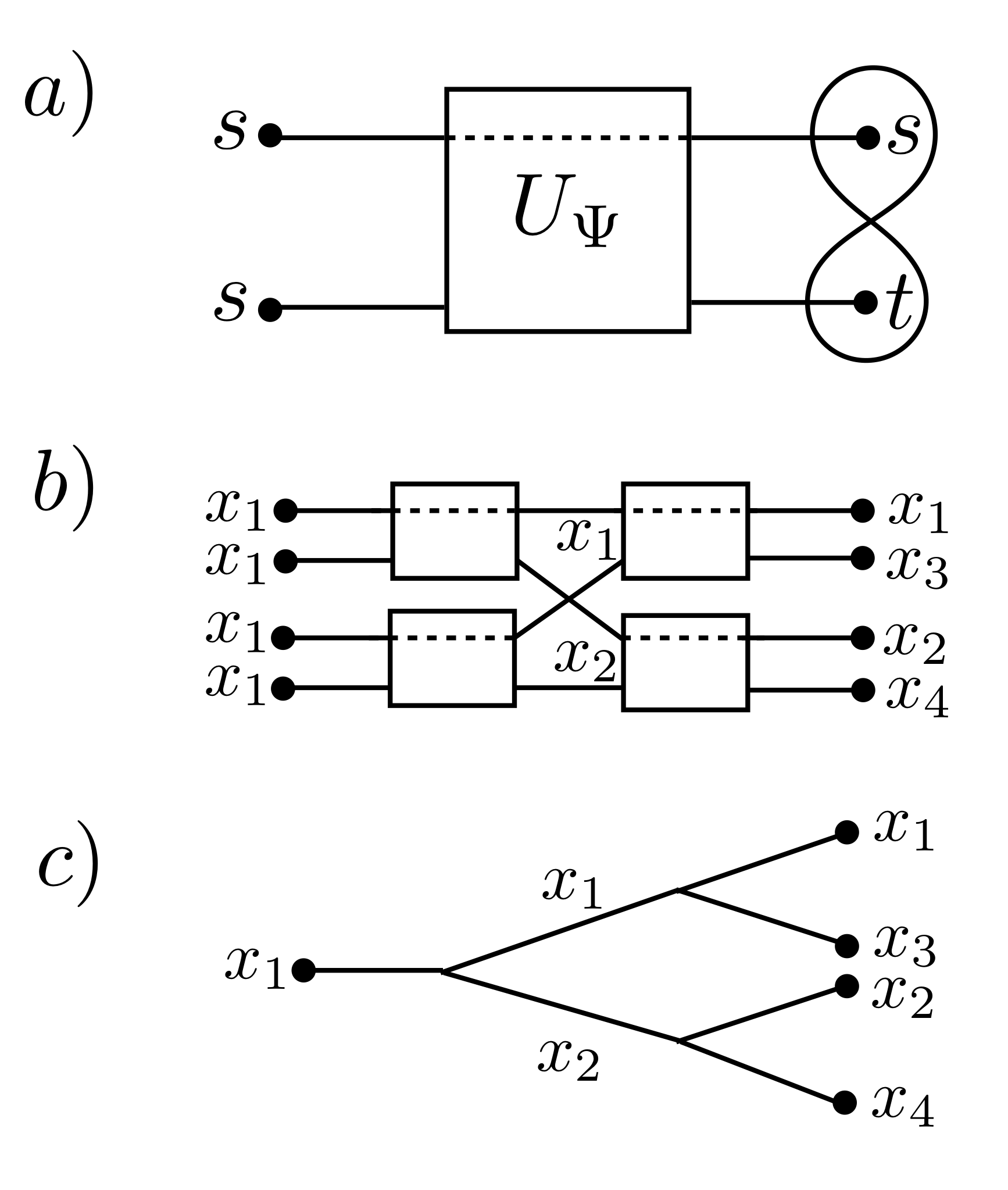}
\caption{a) Graphical representation of the map (\ref{UPsi}), b) example of a circuit for 4 qubits, 
c) associated branching of qubits into four points of space.}
\label{Graphs}
\end{figure}

Secondly, using (\ref{PsiPsiPlus}), one can define the unitary map: 
\begin{align}
U_\Psi := (\mathbb{I} \otimes h^\dagger_R)(\text{CNOT})(\text{H} \otimes \mathbb{I})\,, 
\label{UPsi}
\end{align}
which takes the state $| 0 \rangle_s | 0 \rangle_s$ and generates the maximally entangled 
state (\ref{PsiQubit}) between two particles (qubits) at the space points $s$ and $t$.
The map can be used as a building block for the construction of states 
of multiple particles located at different space points. Such a construction is in the spirit 
of tensor networks, an example of which are quantum circuits. The part a) of Fig. \ref{Graphs} 
contains a graphical representation of (\ref{UPsi}). 

The map, assuming that one of the inputs is an ancilla qubit in the state $|0\rangle$, 
allows us to build a MERA tensor network representing a state of several particles located 
at different positions. However, not necessary the MERA type structures have to be considered. 
An example is shown in the part b) of Fig. \ref{Graphs}, which depicts a circuit generating a state 
of four particles moved to space points $x_1,x_2,x_3,x_4$ from their initial location at 
$x_1$. The part c) of Fig. \ref{Graphs} represents the stages of distributing 
qubits to different space points. 

Thirdly, tensor networks built with the use of blocks (\ref{UPsi}) may correspond 
to the spin network states. However, because Gauss constraint has to be satisfied at the 
nodes of spin networks, the gauge invariance (which is equivalent to the Gauss constraint) 
has to be imposed afterward. One possibility to achieve this is by projecting the obtained 
state onto the spin network basis. 
    
\section{Summary} 

The purpose of this letter was to emphasize the relation between (holonomies of) gauge 
fields and maximally entangled 2-particle states (belonging to the Hilbert space shared by subsystems at two different points in space). The analysis has been performed 
for the case of fundamental representation of the ${\rm SU}(2)$ gauge group but a
generalization to the arbitrary $j$ representation is straightforward. In such a case, the state 
$|\Psi\rangle$ generalizes to $|\Psi\rangle := \frac{1}{\sqrt{2j+1}}\, h^*_{IJ} |I\rangle_s |J\rangle_t$, 
with $I,J = 0,\ldots,2j$. Since an example of the gauge field is the Ashtekar connection 
of the gravitational field, the presented study provides in particular a new perspective on the correspondence 
between the structure of entanglement and the (quantum) geometry of spacetime.  

\section*{Acknowledgements} 

The authors are supported by the Sonata Bis Grant DEC-2017/26/E/ST2/00763 
of the National Science Centre Poland. JM is also supported by the Mobilno\'s\'c Plus Grant 1641/MON/V/2017/0 of the 
Polish Ministry of Science and Higher Education.

\end{document}